\documentclass[3p,times,twocolumn]{elsarticle}
 \biboptions{comma,sort&compress}
 
\usepackage{graphicx}
\usepackage{amsmath}
\usepackage{here}
\usepackage{ecrc}

\volume{00}

\firstpage{1}

\journalname{Nuclear and Particle Physics Proceedings}

\runauth{}

\jid{nppp}

\jnltitlelogo{Nuclear and Particle Physics Proceedings}

\usepackage{amssymb}

\usepackage{lineno}

\usepackage[figuresright]{rotating}

\newcommand*{\myy}{\ensuremath{m_{\gamma \gamma}}}
\newcommand*{\mmm}{\ensuremath{m_{\mu \mu}}}
\newcommand*{\hyy}{\ensuremath{H \rightarrow \gamma \gamma}}
\newcommand*{\hbb}{\ensuremath{H \rightarrow b \bar{b}}}
\newcommand*{\hmm}{\ensuremath{H \rightarrow \mu^{+} \mu^{-}}}
\newcommand*{\hee}{\ensuremath{H \rightarrow e^{+} e^{-}}}
\newcommand*{\ttH}{\ensuremath{t\bar{t}H}}
\newcommand*{\pt}{\ensuremath{p_{T}}}
\newcommand*{\ptv}{\ensuremath{p_{T}^{V}}}
\newcommand*{\GeV}{\text{GeV}}

\begin{document}

\begin{frontmatter}

\title{Latest Studies of the SM Higgs Boson Couplings to Fermions at ATLAS}
 \cortext[cor0]{Talk given at 23rd International Conference in Quantum Chromodynamics (QCD 20),  27 October - 30 October 2020, Montpellier - France} 
 \cortext[cor1]{Copyright 2020 CERN for the benefit of the ATLAS Collaboration. CC-BY-4.0 license.}
 \author[label1]{Yanlin Liu} 
 \author{on behalf of the ATLAS Collaboration}
\address[label1]{University of Michigan}

\pagestyle{myheadings}
\markright{}
\begin{abstract}
This proceeding presents the latest studies on the Yukawa couplings 
of the Standard Model Higgs boson with 139 fb$^{-1}$ data collected using the ATLAS 
detector at a center-of-mass energy of 13 TeV. A first direct probe of $CP$ 
violation in the top-quark Yukawa coupling using events where the Higgs 
boson is produced in association with top quarks ($t\bar{t}H$ and $tH$), and decays into 
two photons (\hyy ) is discussed. The latest results on the Higgs boson production 
in association with a $W$ or $Z$ boson ($VH$) in the $\hbb$ channel 
are depicted as well. Finally, the searches for Higgs boson decays into 
two muons ($\hmm$) and two electrons ($\hee$) are presented.
\end{abstract}

\end{frontmatter}

\section{Introduction}

Since the discovery of the Higgs boson in 2012 \cite{atlashiggs, cmshiggs}, study on its couplings to
the fermions (Yukawa couplings), becomes one important sector
at the Large Hadron Collider (LHC) experiments.
According to the Standard Model (SM), the Yukawa coupling strength is
proportional to the mass of the corresponding fermion. Any experimental deviation
would be a sign of new physics beyond the SM.
From 2015 to 2018, 139 fb$^{-1}$ of $\sqrt{s}$ = 13 TeV proton--proton collision
data was recorded by the ATLAS detector \cite{atlas}.
Based on this amount of dataset, studies on $\ttH$/$tH$ via $\hyy$ \cite{tth} and 
$\hbb$ in the $VH$ production mode \cite{hbbresolved, hbbboosted} (third generation), as well as 
$\hmm$ \cite{hmm} (second generation) and $\hee$ \cite{hee} (first generation) are performed.
Their latest results are reported in the following sections.   

\section{$\ttH$/$tH$ via $\hyy$ Channel}

The Higgs boson production in association with top quarks
provides a direct access to the top-quark Yukawa coupling,
as well as an opportunity to probe the charge conjugation and parity
($CP$) properties of the Higgs boson interactions with the top quarks.
The effective field theory used for this study
is provided by the Higgs Characterization model \cite{hcmodel}.
Within this model,
the term in the effective Lagrangian that describes the top-quark Yukawa coupling is:
\begin{align*}
\cal{L} = &-\frac{m_t}{v} \left\{ \bar{\psi_{t}}\kappa_t \left[\cos(\alpha) + \text{i}\sin(\alpha)\gamma_{\text{5}} \right]\psi_{t} \right\} H
\end{align*}
where $m_{t}$ is the top quark mass, $v$ is the Higgs vacuum expectation value,
$\kappa_t$~($>0$) is the top-quark Yukawa coupling parameter,
and $\alpha$ is the $CP$-mixing angle.
The SM corresponds to a $CP$-even coupling with $\alpha=0$ and $\kappa_t=1$.

Events are selected by requiring two isolated photon candidates with
transverse momenta $\pt$ greater than 35 GeV and 25 GeV,
and at least one jet with $\pt > 25$ GeV  containing a $b$-hadron~($b$-jet),
identified using a $b$-tagging algorithm with an efficiency of 77\% \cite{FTAG-2018-01}.

Selected events are sorted into two $t\bar{t}H$-enriched regions.
The ``Lep'' region, targeting top quark decays in which at least one of the resulting
$W$ bosons decays leptonically, requires events to have at least one
isolated lepton (muon or electron) candidate with $\pt > 15$ GeV.
The ``Had'' region targets hadronic top quark
decays (as well as top quark decays to both hadronically decaying $\tau$-leptons
and unreconstructed leptons) and requires events to have at least two additional
jets with $\pt>25$ GeV and no selected lepton.

In each region, two boosted decision trees (BDTs) are trained:
``Background Rejection BDT'' and ``$CP$ BDT''.
The Background Rejection BDT is trained to separate $t\bar{t}H$-like events from background
which are mainly non-resonant diphoton production processes ($\gamma\gamma$+jets 
and $t\bar{t}\gamma\gamma$).
The $CP$ BDT is trained to separate $CP$-even from $CP$-odd couplings
using $t\bar{t}H$ and $tH$ processes.
The selected events are categorized using partitions
of the two-dimensional BDT space to enhance the analysis sensitivity.
There are 20 categories in total: 12 in the Had region
and 8 in the Lep region.

\begin{figure*}[!htbp]
  \centering
  \includegraphics[width=0.6\textwidth]{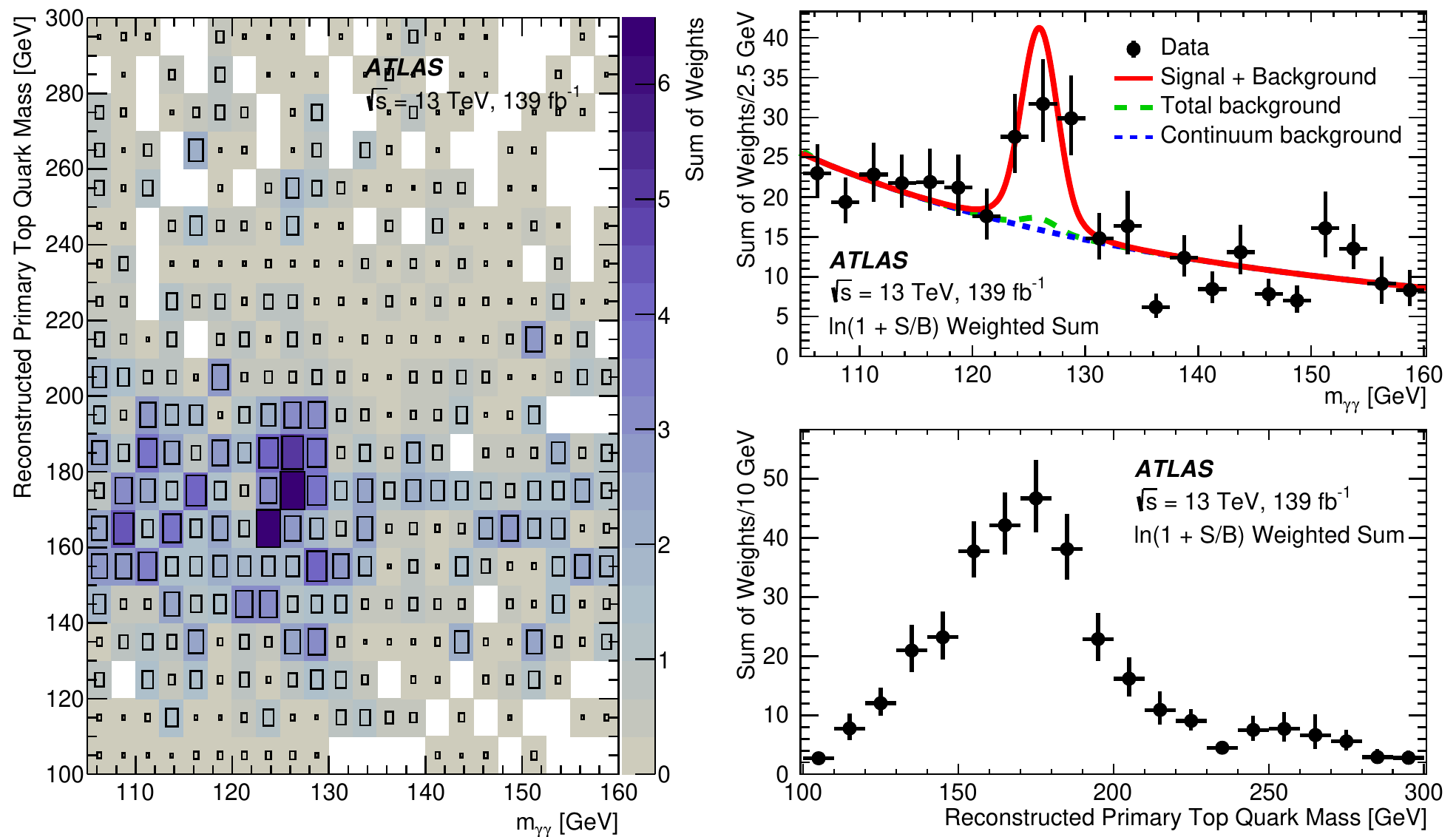}
  \caption{Distribution of reconstructed primary top quark mass versus
  reconstructed Higgs boson mass in the data events.
  The right panels show the projections onto the Higgs boson mass and primary top quark mass axes.
  In the upper panel, the fitted continuum background (blue),
  the total background including non-$t\bar{t}H$/$tH$ Higgs boson production (green),
  and the total fitted signal plus background (red) are shown \cite{tth}.
  }
  \label{fig:mggmt_mc}
\end{figure*}

A simultaneous maximum-likelihood fit is performed to the
diphoton invariant mass ($m_{\gamma\gamma}$) spectra in all the categories 
with all the evaluated systematic uncertainties treated
as nuisance parameters (NPs).
Signal and background shapes are modeled by analytic functions.
Figure \ref{fig:mggmt_mc} shows the distributions of the reconstructed masses for
the diphoton system and primary top quark.
The events
are weighted by $\ln(1 + S/B)$ with $S$ and $B$ being the fitted signal and background
yields in the smallest \myy\ interval containing 90\% of the signal in each category.

Assuming a $CP$-even coupling, the measured rate for $t\bar{t}H$ is
$1.43\;^{+0.33}_{-0.31}\;\text{(stat.)}\;^{+0.21}_{-0.15}\;(\text{sys.})$ times the SM expectation.
The background-only hypothesis is rejected with an observed (expected)
significance of $5.2\sigma$ ($4.4\sigma$).
The $tH$ process is not observed and an upper limit of 12 times the SM expectation
is set on its rate at 95\% confidence level (CL). 

\begin{figure}[!htbp]
  \centering
  \includegraphics[width=0.4\textwidth]{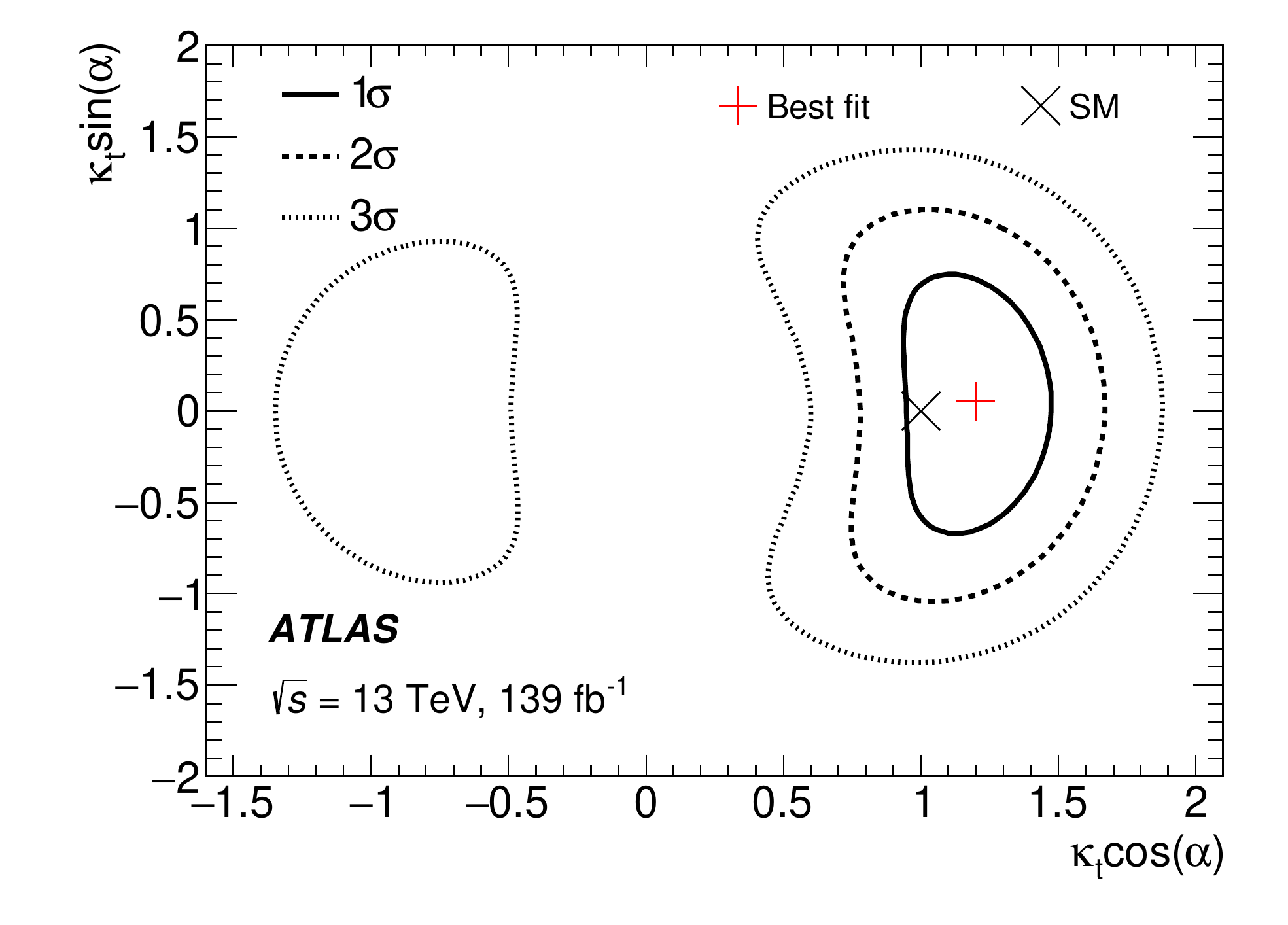}
  \caption{Two-dimensional likelihood contours for $\kappa_{t}\cos(\alpha)$
 and $\kappa_{t}\sin(\alpha)$
with ggF and \hyy\ constrained by the Higgs boson coupling combination \cite{tth}.}
  \label{fig:s3:contours}
\end{figure}

The results of the fit for $\kappa_t \cos(\alpha)$ and $\kappa_t \sin(\alpha)$
are derived with $\hyy$ branching ratio and the Higgs boson coupling to
gluons constrained by the Run 2 Higgs boson coupling combination \cite{hcomb},
and shown as contours in Figure \ref{fig:s3:contours}.
A limit on $\alpha$ is set without prior constraint on
$\kappa_{t}$ in the fit: $|\alpha| > 43^\circ$ is excluded at $95\%$ CL. 
The expected exclusion is $|\alpha| > 63^\circ$
under the $CP$-even hypothesis.
A pure $CP$-odd coupling is excluded at $3.9\sigma$.

\section{$\hbb$ in the $VH$ Production Mode}

$VH$ is the most sensitive production mode for detecting
$\hbb$ decay. The leptonic decay of the vector boson enables 
efficient triggering and a significant reduction of the multi-jet background.
Two analyses are performed: ``Resolved Analysis'' and ``Boosted Analysis'',
which are depicted in the following two subsections.
 
\subsection{Resolved Analysis}

This analysis targets the event topologies containing
exactly a pair of $b$-jets with radius parameter of $R$ = 0.4, 
referred to as small-radius (small-$R$) jets, 
to reconstruct the Higgs boson.
Events are categorized into 0-, 1- and 2-lepton channels 
(referred to as the $n$-lepton channels) depending on the number of
selected electrons and muons, to target the $ZH \rightarrow vv b\bar{b}$, 
$WH \rightarrow \ell \nu b \bar{b}$
and $ZH \rightarrow \ell\ell b\bar{b}$ signatures, respectively.
Events are further split into 2-jet or 3-jet regions, 
where the 3-jet category includes events with one or more untagged jets. 
In the 0- and 1-lepton channels, only one untagged jet is allowed, 
while in the 2-lepton channel any number of untagged jets are accepted in the 3-jet category. 
Since the signal-to-background ratio increases for large 
$\ptv$ (transverse momentum of the vector boson) values,
the selected events are also sorted into two high-\ptv\ regions defined 
as $150\,\GeV<\ptv<250\,\GeV$ and $\ptv>250\,\GeV$.
In the 2-lepton channel, an additional medium-\ptv\ region
with $75\,\GeV<\ptv<150\,\GeV$ is included.

The three $n$-lepton channels, two jet categories and 
two (0-lepton, 1-lepton) or three (2-lepton) \ptv\ regions 
result in a total of 14 analysis regions. 
Each analysis region is further split into one signal region (SR) 
and two control regions (CRs)
using a continuous selection on the angular differences  
between the two $b$-jets as a function of $\ptv$,
resulting in a total of 42 regions. 

The BDT is used to improve the sensitivity of the analysis.
It is trained to discriminate the $VH$ signal from the background processes
in eight regions, obtained by merging some of the 14 analysis regions.
The BDT outputs, evaluated in each signal region, are used as final discriminating variables.

\begin{figure}[!htbp]
  \begin{center}
    \includegraphics[width=0.4\textwidth]{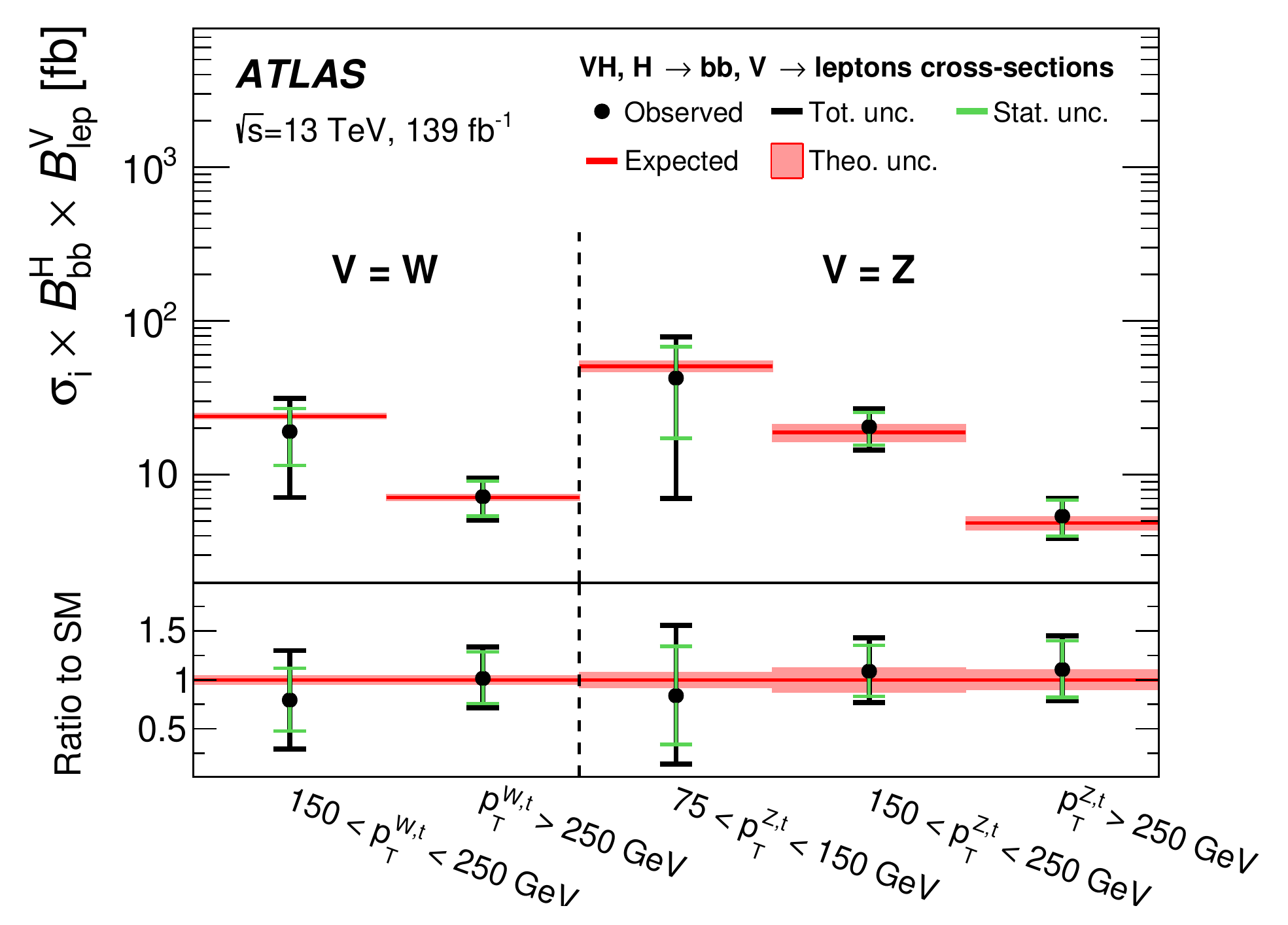}
    \caption{Measured $VH$, $V\to$~leptons cross-sections times the $H\to b\bar{b}$ branching fraction in the reduced STXS scheme \cite{hbbresolved}. }
    \label{fig:Xsections}
  \end{center}
\end{figure}

A binned likelihood fit is performed to the BDT outputs in all the categories.
The effects of systematic uncertainties enter the likelihood as NPs.
The normalizations of the dominant background $V$+HF (heavy flavor)
 and $t\bar{t}$ are left
unconstrained in the likelihood, and determined by the fit. 
The $WH$ and $ZH$ production modes reject the background-only hypothesis 
with observed (expected) significances of 4.0$\sigma$ (4.1$\sigma$) 
and 5.3$\sigma$ (5.1$\sigma$), respectively.
The total observed (expected) significance for $VH$ production mode
is 6.7$\sigma$ (6.7$\sigma$).
Figure \ref{fig:Xsections} shows the measured $VH$ cross-sections times
the $\hbb$ and $V\rightarrow$ leptons branching fractions,
$\sigma\times B$, together with the SM predictions in the reduced 
simplified template cross-section (STXS) \cite{hbbstxs} regions.
The measurements are all consistent with the SM predictions 
with relative uncertainties varying from 30\% in the highest 
\ptv\ region to 85\% in the lowest \ptv\ region. 
The data statistical uncertainty is the largest single uncertainty,
and the major systematic uncertainties arise from 
the background modelling, $b$-tagging correction factors 
as well as jet energy scale calibration and resolution.

\subsection{Boosted Analysis}

This analysis aims for high $\pt$ Higgs boson regime, where the Higgs boson
is reconstructed as a single large-$R$ jet ($J$) with $R$ = 1.0 and
at least two track-jets associated.
The large-$R$ jets are required to have $\pt > 250$ GeV,
mass $m_\textrm{J}>50$ GeV and pseudorapidity $|\eta| < 2.0$.
The leading two track-jets associated with $J$ are required to be $b$-tagged.

Events are categorized into 0-, 1- and 2-lepton channels
depending on the number of selected electrons and muons.
10 SRs and 4 CRs are defined based on $\ptv$
($250<\ptv<400\,\GeV$ or
$\ptv > 400\,\GeV$) and
the number of small-$R$ jets not matched to $J$.

\begin{figure}[!htbp]
  \centering
\includegraphics[width=0.4\textwidth]{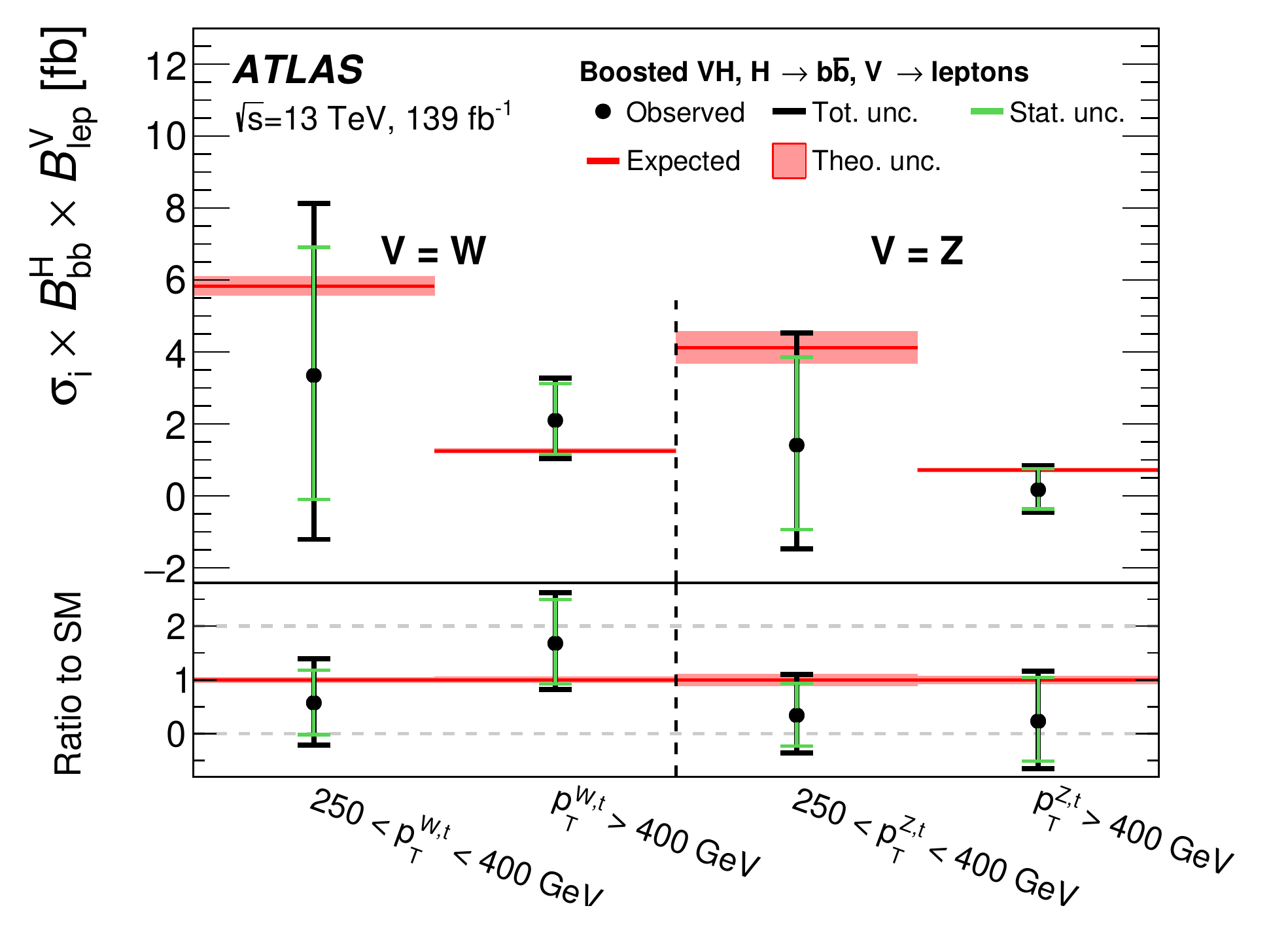}
\caption{Measured
  $VH$ reduced stage-1.2 simplified
  template cross-sections times the $\hbb$ and
  $V \rightarrow$~leptons branching fractions \cite{hbbboosted}.}
  \label{fig:STXSBinXSPlot}
\end{figure}

The results are obtained from a binned maximum-profile-likelihood
fit to the observed $m_\textrm{J}$ distributions in all the SRs and CRs.
The normalizations of the dominant background $V$+jets and $t\bar{t}$
are left unconstrained in the likelihood, and determined by the fit.
The observed (expected) significance is 2.1$\sigma$ (2.7$\sigma$) for $VH$, $\hbb$ process.
Figure~\ref{fig:STXSBinXSPlot} shows the measured $VH$ cross-section 
times branching fractions $\sigma \times B$ in each STXS bin for the reduced
scheme, together with the SM predictions.
For these results, the largest uncertainty arises from data statistics.
The major systematic uncertainty is related to the large-R jet calibration, 
particularly the $m_\textrm{J}$ resolution. 

\section{$\hmm$}

The $\hmm$ decay offers the best opportunity to
measure the Higgs boson interactions with a second-generation fermion at the LHC.
This analysis selects events with two isolated muons with opposite charge and
the leading muon is required to have $\pt>27$ GeV.
To increase the signal sensitivity, the selected 
events are classified into 20 mutually exclusive categories
based on the event topology and BDT discriminants.

A category enriched in $\ttH$ events is defined in order to target the 
dileptonic or semileptonic decay of the $t \bar{t}$ system.
Events are considered for this category if there is at least one lepton ($e$ or $\mu$)
with $\pt > 15\,\GeV$ in addition to the opposite-sign muon pair
and at least one $b$-jet 
identified using a $b$-tagging algorithm with an efficiency of 85\%.
A BDT is trained using simulated $\ttH, \hmm$ events as signal  
and simulated events from all SM background processes as background.
A selection is applied to the BDT score to define one $\ttH$-enriched category.

Events not selected in the $\ttH$ category are considered for the
$VH$-enriched categories. The $VH$ categories target signal events
where the Higgs boson is produced in association with a leptonically
decaying vector boson.
Events are required to have at least one additional isolated lepton. 
Two BDTs are trained, separately for the three-lepton and four-lepton events, 
to discriminate between the simulated signal and background events.
Based on the BDT scores, two categories are defined for the three-lepton
events and one category is defined for the four-lepton events.

The events not selected in the $\ttH$ or $VH$ categories, 
are further classified into three
channels according to the jet multiplicity: 
0-jet, 1-jet and 2-jet, where the last includes events with two or more jets.
In the 2-jet channel, a BDT is trained to disentangle signal
events produced by VBF, used as signal sample in the training, from background events.
Four VBF categories are defined based on this BDT classifier.
The remaining events are classified by three other BDTs (split by jet multiplicity),
which are trained with both the $\hmm$ ggF
and VBF production MC samples as signal.
Four categories are defined in each of the three BDTs, resulting in
twelve ggF categories in total.

\begin{figure}
  \centering
  \includegraphics[width=0.4\textwidth]{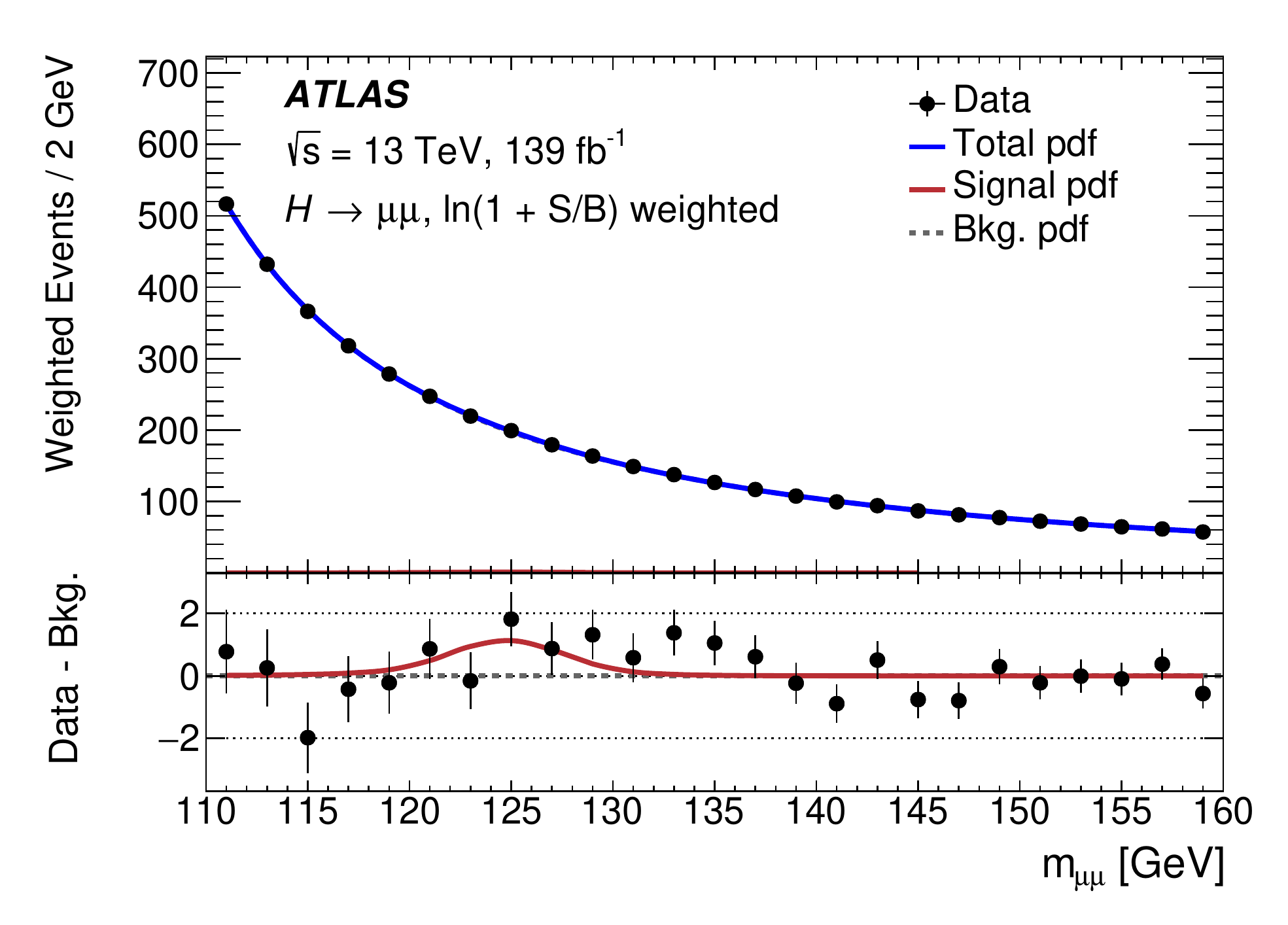}
  \label{fig:InclusiveW}
  \caption{ Dimuon invariant mass spectrum in all the analysis
    categories observed in data. 
    Events and pdfs are 
    weighted by $\ln(1 + S/B)$, where $S$ are the observed signal yields 
    and $B$ are the background yields
    derived from the fit to data in the $\mmm = 120$ -- $130\,\GeV$ window \cite{hmm}.
    }
  \label{fig:Inclusive}
\end{figure}

\begin{figure}
  \centering
  \includegraphics[width=0.4\textwidth]{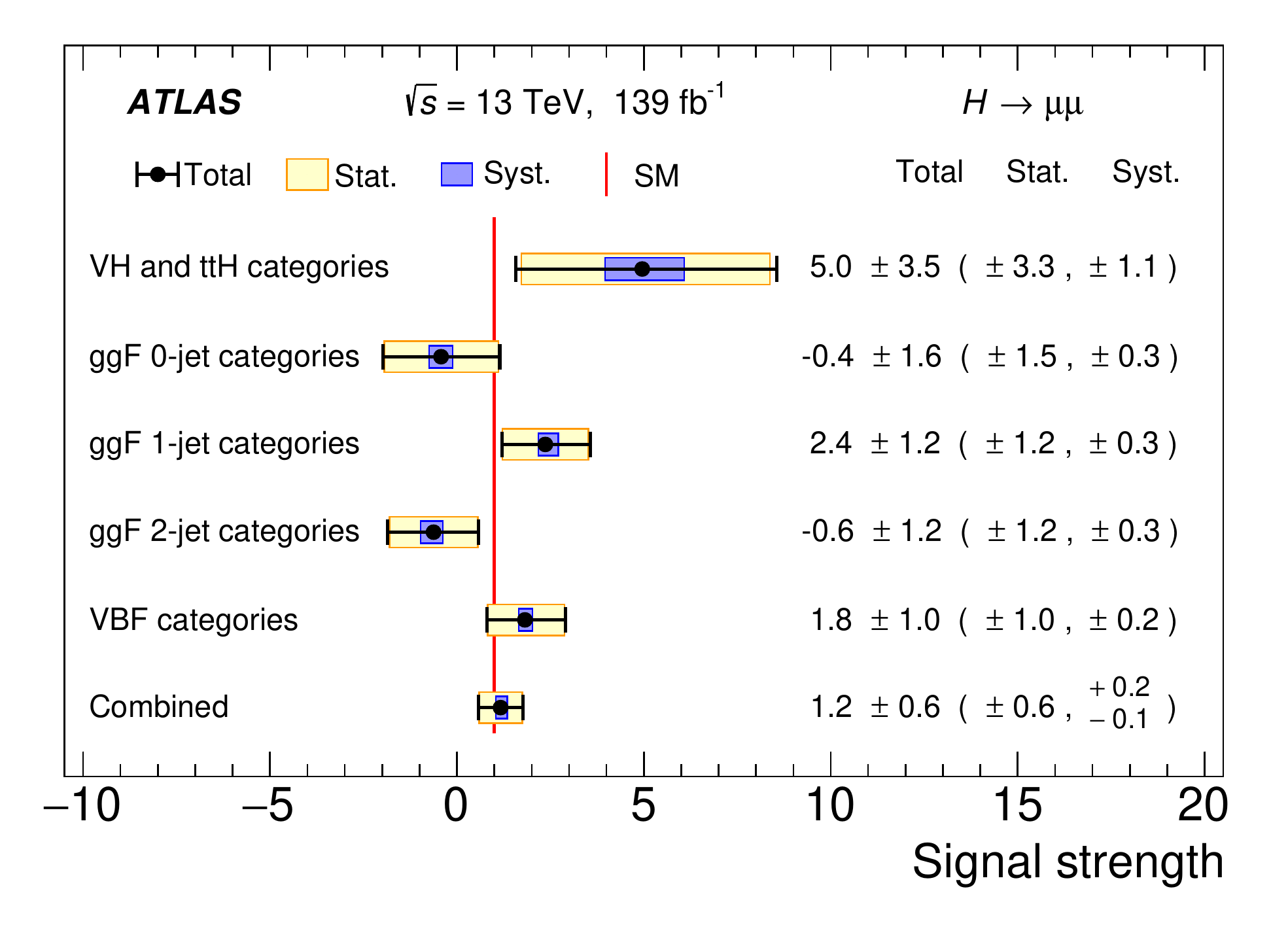}
  \caption{ The best-fit values of the signal strength parameters for the five major groups of categories  together with the combined value \cite{hmm}.}
  \label{fig:5categories}
\end{figure}

The signal yield is obtained by a simultaneous binned maximum-likelihood 
fit to the dimuon invariant mass \mmm\ distributions of the 20 
categories in the range 110 -- 160\,\GeV .
Analytic models are used in the fit to describe the \mmm\ 
distributions for both the signal 
and background processes.
Figure \ref{fig:Inclusive} shows the \mmm\ spectra for 
all the analysis categories after the signal-plus-background fit.
The best-fit value of the signal strength parameter,
defined as the ratio of the observed signal yield to the one
expected in the SM, is
$\mu = 1.2 \pm 0.6$, corresponding to an observed (expected)
significance of $2.0 \sigma$ ($1.7 \sigma$) with respect to the
hypothesis of no $\hmm$ signal.
The best-fit values of the signal
strength parameters for the five major groups of categories 
($\ttH$+$VH$, ggF 0-jet, 1-jet, 2-jet, and VBF)
together with the combined value
are shown in Figure \ref{fig:5categories}.

\section{$\hee$}

This is the first ATLAS search for $\hee$ process, which has very small
branching fraction (around $5 \times 10^{-9}$).
Events are selected by requiring exactly two isolated electrons with opposite charge,
and the leading electron needs to have $\pt>27$ GeV.

To improve the overall sensitivity of this search,
selected events are divided into seven categories.
First, a category enriched in
events from VBF production is defined by selecting those containing two jets with
pseudorapidities of opposite signs, a pseudorapidity separation
$|\Delta \eta_{jj}|>3$ and a dijet invariant mass $m_{jj}>500$ GeV.
Events that fail to meet the criteria of the VBF
category are classified as ``Central'' if the pseudorapidities of
both leptons are $|\eta^{e}|<1$ or as ``Non-central'' otherwise. For each of these two
regions, three categories are defined based on the dielectron transverse momentum 
$\pt^{ee}$: ``Low $\pt^{ee}$'' ( $\pt^{ee} \le 15$ GeV), ``Mid
$\pt^{ee}$'' ($15<\pt^{ee} \le 50$ GeV), and ``High
$\pt^{ee}$'' ($\pt^{ee} >50$ GeV).

\begin{figure}[!htbp]
  \centering
  \includegraphics[width=0.4\textwidth]{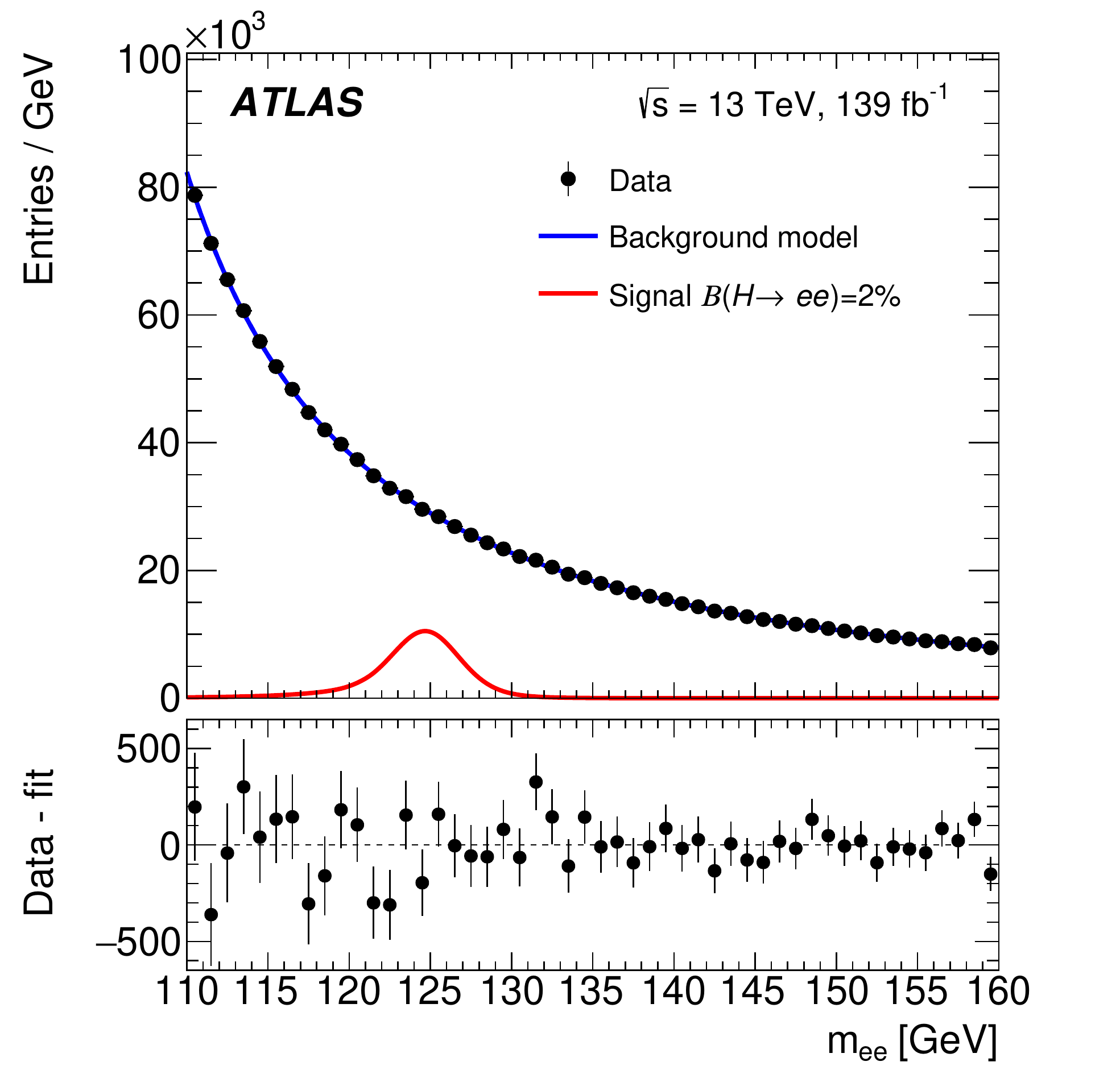}

  \caption{Dielectron invariant mass for all categories summed together
    compared with the background-only model. The
    signal parameterisations with branching fractions set to
    $\mathcal{B}(\hee) = 2\%$ are
    also shown (red line). The bottom panels show the difference
    between data and the background-only fit \cite{hee}.
}
  \label{fig:heemass}
\end{figure}

The signal yield is obtained by a simultaneous binned maximum-likelihood
fit to the dielectron invariant mass spectra of the seven
categories in the range 110 -- 160\,\GeV .
The data and expectation for all categories summed
together are shown in Figure~\ref{fig:heemass}.
No evidence of the $H\to ee$ process is found.
The observed (expected) upper limit on the branching fraction
is set as $3.6 \times 10^{-4}$ ($3.5 \times 10^{-4}$) at the
95\% CL.

\section{Summary}

The latest studies of the SM Higgs boson couplings to the fermions
with 139 fb$^{-1}$ data are presented. All the measured results
are consistent with the SM predictions, and they represent significant
improvements on sensitivity and precision comparing with previous publications.

\end{document}